\documentstyle[twocolumn,aps,epsfig]{revtex}

\begin{document}
\title{Probing exotic spin correlations by Muon Spin depolarization measurements
with applications to spin glass dynamics.}
\author{$^{1}$Amit Keren, $^{1}$Galina Bazalitsky, $^{2}$Ian Campbell, and $^{3}$%
James S. Lord}

\address{
$^{1}$Physics Department, Technion-Israel Institute of Technology, 
Haifa 32000, Israel.\\
$^{2}$Laboratoire des Verres, Universit\'{e} Montpellier II, 91405
Montpellier, France. \\
$^{3}$Rutherford Appleton Laboratory, Chilton Didcot, Oxfordshire OX11 0QX, U.K.}

\address{{\rm (Received: )}}
\address{\mbox{ }}
\address{\vspace{4mm}\parbox{14cm}{\rm \mbox{ }\mbox{ }\\ We develop a new method 
to probe the local spin dynamic autocorrelation
function, using magnetic field dependent muon depolarization measurements.
We apply this method to $\mu $SR experiments in the dilute Heisenberg spin
glass {\it Ag}Mn($p$ at. \%) at $T>T_{g}$ where the correlations of the Mn
local magnetic moment are strongly non-exponential. Our results clearly
indicate that the dynamics of this spin glass cannot be described by a
distribution of correlation times. Therefore, we analyze the data assuming a
local spin correlation function which is the product of a power law times a
cutoff function. The concentration and temperature dependence of the
parameters of this function are determined. Our major conclusion is that in
the temperature region close to $T_{g}$ the correlation function is
dominated by an algebraic relaxation term.\\}}
\address{\mbox{ }}
\address{\vspace{2mm}\parbox{14cm}{\rm PACS numbers: }}
\maketitle

\section{Introduction}

\label{Intoduction}

In the vicinity of a second order magnetic phase transition, the spin-spin
dynamical autocorrelation function 
\begin{equation}
\left\langle {\bf S}_{i}(t){\bf S}_{i}(0)\right\rangle \equiv q(t),
\label{DefOfq}
\end{equation}
must take the general form

\begin{equation}
q(t)\sim t^{-x}f(t/\tau )  \label{SpinSpinCorr}
\end{equation}
\cite{HohenbergRMP77}. In these equations $i$ is a site index, $\left\langle
{}\right\rangle $ is a thermal and sample average, $\tau $ is a time scale
that limits the upper range of the decay, $x$ can be defined in terms of
the static and dynamic critical exponents, and $f$ is a ``cutoff'' function
which can take different forms. In simple systems, such as isolated spins,
the power law part does not exist ($x=0$), and $f$ is exponential, making $%
q(t)$ exponential. In systems with interacting spins $q(t)$ can be
non-exponential. For example, in a pure ferromagnet close to $T_{c}$, $q(t)$
is given by Eq.~\ref{SpinSpinCorr} with $x>0$ and exponential $f$. In this
case, the correlation function of all individual spins is the same as the
global one by definition. In complex systems, with physical properties such
as time scale invariance \cite{TimeScale} or hierarchical relaxation \cite
{PalmerPRL84} $q(t)$ can be {\em strongly} non-exponential. An important
example is spin glasses with frozen-in disorder, where the local
environments of individual spins are not identical. In this case numerical
simulations show that the global correlation function also takes the form of
Eq.~\ref{SpinSpinCorr} but $f$ is a ``stretched exponential'' \cite
{OgielskiPRB85,FranzesePRE99}. In addition, these simulations show that Eq.~%
\ref{SpinSpinCorr} is valid even locally but $f$ varies from spin to spin.
Therefore, they have been analyzed with more complex cutoff functions, such
as a sum of exponentials \cite{TakayamaPhyA94}, or a sum of stretched
exponentials \cite{GlotzerPRE98}.

Despite this theoretical understanding of the correlation function, the
complex nature of $q(t)$ is frequently ignored in the analysis of
experimental systems and they are interpreted in terms of a sum of
exponential relaxation rates: some spins are relaxing fast and some slowly.
Indeed, the Laplace transform of any relaxation function $q(t)$ can always
be defined in terms of a spectrum of relaxation times. However, the
relaxation times of the distribution are those of the modes of the system, 
{\it not} those of the individual spins. (To take a familiar analogy, in a
pure crystal there is a spectrum of phonon modes but any atom is vibrating
in the precisely the same complex way as any other atom).

In this work we avoid the assumption of a distribution of correlation times
and determine experimentally $q(t)$ using the alternative hypothesis of a
unique non-exponential $q(t)$. For reasons which are described in Sec.~\ref
{Experiment} our technique of choice is Muon Spin Relaxation [$\mu $SR].
Muons, are implanted in a sample initially fully polarized, and are
depolarized in the neighborhood of local magnetic moments. The form of the
muon depolarization is a signature of $q(t)$. The practical difficulty is to
establish in the general case an inversion scheme that can take one
unambiguously from the muon depolarization pattern back to the correlation
function. Accordingly, our main objectives in this work are: (I) to develop
(in Sec.~\ref{Analysis}) a method by which one can determine the functional
form of $q(t)$ by measuring both the field and time dependence polarization $%
P(H,t)$ of the muon; (II) to apply this method to a real Heisenberg spin glass
and to determine $x$ and $\tau $ (see Sec.~\ref{Results}).

This work is a continuation of our previous publication where we have
demonstrated the existence of an unusual correlation function in the spin
glass {\it Ag}Mn(0.5~at.~\%) at $T>T_{g}$ at a single temperature \cite
{KerenPRL96}. There it was shown that at {\em high fields} ($>$120~G) and 
{\em late times}, the muon polarization obeys the scaling relation 
\begin{equation}
P(H,t)=P(t/H^{1-x}),  \label{ScalingRelations}
\end{equation}
and this relation was traced back to the correlation function given by Eq.~\ref
{SpinSpinCorr}. However, an interpretation of the parameters $x$ or $\tau $
was not provided, and no attempt was made to discuss the singularity of $%
q(t) $ at $t=0$. Here we extend Ref. \cite{KerenPRL96} in four different
directions: (I) we address the $t=0$ singularity, (II) we apply lower fields
where $P(H,t)$ is most sensitive to $f$, (III) we vary the temperatures, and
(IV) we examine three different samples of {\it Ag}Mn($p$~at.~\%) with $%
p=0.1 $, $0.3$, and $0.5$ where $T_{g}=0.51$, $1.45$, and $2.80$~K
respectively. This allows us to estimate the various parameters of $q(t)$ as
a function of $p$ and $T$. Our most important finding is that $q(t)$ depends
on temperature mainly through $x(T)$ in these samples.

\begin{figure}
\unitlength=1mm
\begin{picture}(150,95)
\put(5,0){\epsfig{file=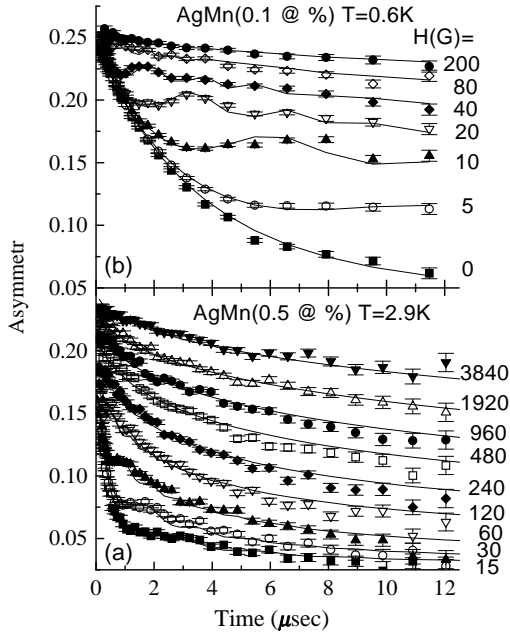,height=100mm,width=70mm}}
\end{picture}
\caption{Asymmetry vs time in two samples for various longitudinal magnetic
fields. The solid lines are fit to the model decribed in the text.}
\label{Fitting}
\end{figure}

\section{Experimental aspects and raw data}

\label{Experiment}

As demonstrated in Ref.~\cite{KerenPRL96} and derived again in Sec.~\ref
{Analysis} a clear evidence for an exotic correlation function is the
scaling relation of Eq.~\ref{ScalingRelations}. In order to find these type
of scaling experimentally from the raw muon spectra with good certainty, one
must vary the external field over two orders of magnitude. The $\mu $SR
technique is specially adapted for this task since the signal intensity is
independent of $H$, and one can vary the magnetic field considerably, the
upper limit being given by the fields available in the experimental set ups,
and the lower limit being related to the range of applicability of the
analysis method. In addition, experiment shows that the typical time scales
are of the order of a microsecond, which falls between the ranges covered
conveniently by neutron techniques and by conventional susceptibility
techniques.

Our $\mu $SR experiments were performed in ISIS where magnetic fields up to $%
4$~kG are available. In these experiments we measure the time and field
dependent asymmetry $A(H,t)$ of the decay positrons in the direction
parallel and anti parallel to the initial muon polarization. This asymmetry
is proportional to $P(H,t)$ where $H$ is applied in the direction of the
initial polarization, and $P(H,0)=1$; for more details see Ref. \cite
{KerenPRL96}. The fields applied for each sample obey the condition $\mu
_{eff}H<T_{g}$, where $\mu _{eff}$ is the magnetic moment of the Mn ions.
This condition ensures that the impact of the fields on the local moment
correlation function is minimal. Standard frequency dependent susceptibility
measurements on spin glasses show that a magnetic field has negligible
effect on the spin glass relaxation, even at quite low frequencies
corresponding to times of the order of a millisecond; as the muSR technique
is intrinsically limited to time scales of the order of a microsecond or
less, we can safely ignore any direct effects of magnetic fields on the
local moment relaxation.

\begin{figure}
\unitlength=1mm
\begin{picture}(100,70)
\put(5,0){\epsfig{file=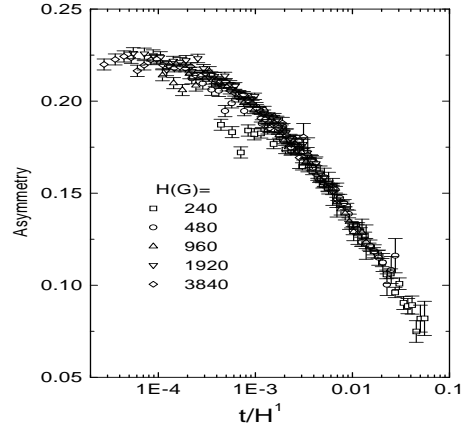,height=70mm,width=70mm}}
\end{picture}
\caption{The asymmetry from Fig.~1a plotted as a function of $t/H^{1}$ for
fields higher than 240~G. This plot demonstrates the experimental validity
of Eq.~\ref{ScalingRelations}.}
\label{Scaling}
\end{figure}

We have used {\it Ag}Mn since Ag nuclear moments are negligible and there is
no need to worry about background relaxation from Ag nuclei. Though dilute,
the Mn nuclear moments are large and it turns out that the Mn nuclear term
can influence the muon depolarization signal in zero applied field. The
samples were prepared by melting in a sealed quartz tube, followed by strong
cold work. The samples were transformed from buttons to sheets about 0.2 mm
thick and were then annealed for a week under vacuum. This technique is
standard for ensuring complete homogenization. The glass transition
temperatures were determined using a standard ac-susceptibility method. For
the $p=0.3$ and $p=0.1$ concentrations we were assisted by the dilution
refrigerator of C. Paulsen in Grenoble.

Representative $A(H,t)$ measurements in the $p=0.5$ and $0.1$ samples at $%
T=2.9$ and $0.6$ K respectively, and at various fields, are shown in Fig.~%
\ref{Fitting} panels (a) and (b). Clearly, $H$ affects the muon polarization
very strongly over more than two orders of magnitude. The larger $H$ the
weaker the relaxation. This well known effect is called decoupling. The
scaling relations of Eq.~\ref{ScalingRelations} are shown to hold with $%
x\rightarrow 0$ at {\em high fields} and {\em late times }for the $p=0.5$
sample in Fig.~\ref{Scaling}. Our aim is to account for the rate of
decoupling, namely, the scaling relations, quantitatively.

\section{Analysis}

\label{Analysis}

First we would like to emphasize the need for a new type of analysis for our
data. In the appendix we show that the muon polarization is determined by
the field-field dynamical auto correlation function 
\[
\Phi (t)\equiv \gamma _{\mu }^{2}\left\langle {\bf B}_{\bot }^{d}(t)\cdot 
{\bf B}_{\bot }^{d}(0)\right\rangle 
\]
(rather than by $q$) where ${\bf B}_{\bot }^{d}$ is the dynamic magnetic
field at the muon site perpendicular to the external field ${\bf H}$, and $%
\gamma _{\mu }$ $=85.16$~MHz/kG is the muon gyromagnetic ratio. In the {\em %
high fields }and{\em \ late time} limit we arrive at the relation 
\begin{equation}
P(H,t)=P\left[ t/T_{1}(H)\right]  \label{PHtSimple}
\end{equation}
where 
\begin{equation}
\frac{1}{T_{1}(H)}=\int_{0}^{\infty }\Phi (t^{\prime })\cos (\gamma
Ht^{\prime })dt^{\prime }.  \label{T1Simple}
\end{equation}
Since in a spin glass there are no cross correlations between spins one
expects $\Phi (t)\propto q(t)$. Therefore, for an exponential correlation
function $q(t)$ with a single correlation time we write 
\begin{equation}
\Phi (t)=2\Delta ^{2}\exp (-t/\tau )  \label{Trivial}
\end{equation}
where 
\[
2\Delta ^{2}=\gamma _{\mu }^{2}\left\langle {\bf B}_{\bot }^{2}\right\rangle
. 
\]
In this case the muon relaxation rate is given by the known expression 
\[
\frac{1}{T_{1}(H)}=\frac{2\Delta ^{2}\tau }{1+\left( \gamma _{\mu }\tau
H\right) ^{2}}. 
\]

However, in a spin glass, the muon could stop in a variety of environments
and can experience different instantaneous fields or correlation times. If
we allow for a distribution of $\Delta $ and/or $\tau $ we can obtain an
average polarization $\overline{P}$ by 
\[
\overline{P}(H,t)=\int \int \rho (\Delta ,\tau )P\left( \frac{2\Delta
^{2}\tau t}{1+\left( \gamma _{\mu }\tau H\right) ^{2}}\right) d\Delta d\tau 
\]
Nevertheless, if $\gamma _{\mu }H\gg 1/\tau ^{\min }$, where $\tau ^{\min }$
is the shortest correlation time in the distribution we expect 
\[
P(H,t)=P\left( t/H^{2}\right) 
\]
which is in contrast to the experimental observation depicted in Fig.~\ref
{Scaling}. This brings us to one of the important conclusions of our
experiments: a distribution of correlation times could not explain our data.

The second problem with the standard analysis method emerges from the
``wiggles'' seen in the data at {\em early times }and{\em \ low fields}.
These could not be accounted for by Eq.~\ref{PHtSimple} which does not have
oscillating terms. The source of the oscillations are those muons for which
the transverse magnetic field has not relaxed abruptly during the time of
one rotation. These muons will oscillate around the vector sum of both
internal and external fields at a frequency $\omega =\gamma _{\mu }\sqrt{%
H^{2}+B_{\bot }^{d}}$. Since at $T>T_{g}$ there is a distribution of $B^{d}$
with zero average, the contribution of the dynamic field will show up as a
relaxation of the wiggles, while the frequency of the wiggles will be at $%
\omega =\gamma _{\mu }H$. Roughly, if there is a cutoff time $\tau $, then
by the time $t>\tau $ abrupt field relaxation occurs for all muons.
Therefore, wiggle pattern will not be seen for longer times. As a result,
the observation of a ``wiggle'' pattern up to time $t$ is the proof of local
moments which do not relax abruptly within this time scale, and so gives a
model independent estimate of $\tau $. The fit procedure described below
provides a quantitative method to include this type of observation within
the global analysis of the field dependent muon depolarization.

Having demonstrated that neither the field-time scaling relations nor the
wiggles in the data could be accounted for with standard analysis methods we
turn to describe our alternative approach. If one does not take the {\em %
long time} limit that led to Eqs.~\ref{PHtSimple} and \ref{T1Simple} one
finds (see appendix) that the relation between $P(H,t)$ and $\Phi (t)$ is
via the expression 
\begin{equation}
P(H,t)=P_{0}\exp \left( -\Gamma (H,t)t\right) ,  \label{DefOfGamma}
\end{equation}
where 
\begin{equation}
\Gamma (H,t)t=\int_{0}^{t}(t-t^{\prime })\Phi (t^{\prime })\cos (\gamma
_{\mu }Ht^{\prime })dt^{\prime }.
\end{equation}
This expression can produce the oscillations seen in the data (for the
exponential case given by Eq.~\ref{Trivial} see \cite{KerenPRB94}).
Motivated by the success of Eq.~\ref{SpinSpinCorr} to account for the
preliminary data of Ref.~\cite{KerenPRL96} we examine here the correlation
function 
\begin{equation}
\Phi (t)=2\Delta ^{2}\frac{\tau _{e}^{x}}{(t+\tau _{e})^{x}}\exp (-t/\tau
_{l})  \label{PartDefOfq}
\end{equation}
where $\tau _{e}$ and $\tau _{l}$ are an early and late time cutoff
respectively. This $\Phi (t)$ is properly normalized at $t=0$. From neutron
spin echo data \cite{MezeiJMMM79}, $\tau _{e}$ is of the order of $1$%
~picosecond, and approximately independent of temperature. Therefore, we are
safe to assume that the time at which the muon polarization is first
measured (0.1~$\mu $sec) is much longer than $\tau _{e}$. In other words,
our experiment is in the limit of $t\gg \tau _{e}$, and we can write 
\[
\Phi (t)=2C\iota (t) 
\]
where 
\[
C=\Delta ^{2}\tau _{e}^{x} 
\]
and 
\begin{equation}
\iota (t)=t^{-x}\exp (-t/\tau _{l}).  \label{Itao}
\end{equation}
Now we express 
\begin{equation}
\Gamma (H,t)t=2\Delta ^{2}\tau _{e}^{x}\int_{0}^{t}(t-t^{\prime })\iota
(t^{\prime })\cos (\gamma Ht^{\prime })dt^{\prime }.  \label{GammaOfTime}
\end{equation}

\begin{figure}
\unitlength=1mm
\begin{picture}(150,95)
\put(5,0){\epsfig{file=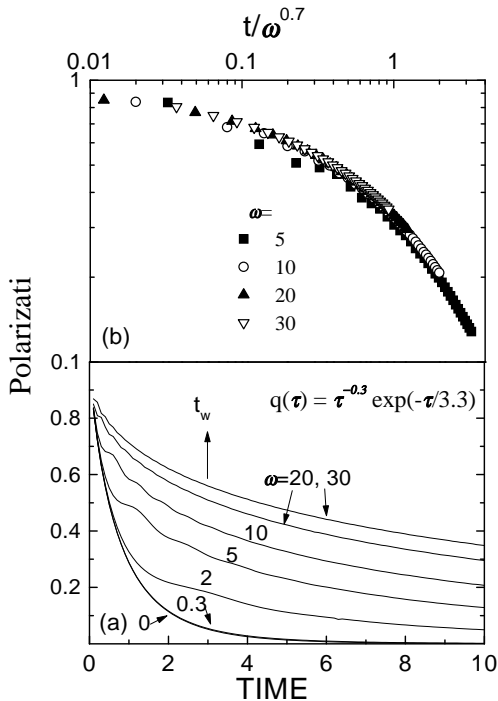,height=100mm,width=70mm}}
\end{picture}
\caption{(a) The polarization generated using Eqs.~\ref{Itao}, \ref{SGKeren}%
, and \ref{DefofGamma}, for various frequencies $\protect\omega =\protect%
\gamma _{\mu}H$. (b) Demonstrating the theoretical validity of Eq.~\ref
{ScalingRelations} for high enough values of $\protect\omega =\protect\gamma
_{\protect\mu }H$ and late times.}
\label{Numerics}
\end{figure}

To be rigorous we should allow for a distribution of all the parameters in
the correlation function. This, of course, will lead us nowhere. Therefore,
we assume only a distribution of $\Delta $ and so assume a unique form of $%
q(t)$ for all local spins. This assumption appears physically reasonable and
as we shall see it allows us to account for our data successfully. We
note that all $\mu $SR data to date were analyzed using this assumption,
and, in particular, the following distribution \cite{UemuraPRB85} was chosen 
\begin{equation}
\rho (\Delta )=\sqrt{\frac{1}{2\pi }}\frac{\Delta ^{\ast }}{\Delta ^{2}}\exp
\left( -\frac{1}{8}\left[ \frac{\Delta ^{\ast }}{\Delta }\right] ^{2}\right)
.  \label{DisDelta}
\end{equation}
As before the average muon polarization $\overline{P}$ is given by 
\[
\overline{P}(H,t)=P_{0}\int \rho (\Delta )\exp \left( -\Gamma (H,t)t\right)
d\Delta 
\]
which leads to 
\begin{equation}
\overline{P}(H,t)=P_{0}\exp \left( -\left[ \Gamma ^{\ast }(H,t)t\right]
^{1/2}\right)  \label{SGKeren}
\end{equation}
where 
\begin{equation}
\Gamma ^{\ast }(H,t)t=2C^{\ast }\int_{0}^{t}(t-t^{\prime })\iota (t^{\prime
})\cos (\gamma Ht^{\prime })dt^{\prime },  \label{DefofGamma}
\end{equation}
and 
\begin{equation}
C^{\ast }=(\Delta ^{\ast })^{2}\tau _{e}^{x}.  \label{CStar}
\end{equation}
The difference between $P$ before averaging over $\Delta $, and $\overline{P}
$ (after the average) is fundamental and is manifested in the appearance of
a power $1/2$ in the exponent of Eq.~\ref{SGKeren}. In fact, the reason for
choosing $\rho (\Delta )$ given by Eq.~\ref{DisDelta} is that experimental
data agrees well with Eq.~\ref{SGKeren}. In contrast, the difference between 
$\Gamma $ and $\Gamma ^{\ast }$ is minor; $C$ is simply replaced by $C^{\ast
}$, namely, $\Delta $ is replaced by $\Delta ^{\ast }$. In reality we do not
use Eq.~\ref{SGKeren} but rather fit our data to 
\begin{equation}
\overline{P}(H,t)=P_{0}\exp \left( -\left[ \Gamma ^{\ast }(H,t)t\right]
^{\beta }\right)  \label{FinalPolarization}
\end{equation}
with $\beta \sim 1/2$ a global parameter for all samples and temperatures.

It is instructive to examine Eq. \ref{FinalPolarization} and \ref{DefofGamma}
numerically for a particular example of $\iota (t)$. The integral of Eq.~\ref
{DefofGamma} is well defined and could be evaluated by the ``improper
integration'' method \cite{NumericalRecepes}. In Fig.~\ref{Numerics}a we depict the
numerical $P(\omega ,t)$ vs. $t$ for various values of $\omega =\gamma _{p}H$
where the parameters of the correlation function are $C^{\ast }=1$, $x=0.3$, 
$\tau _{l}=3.3$, and in Eq.~\ref{FinalPolarization} $\beta =0.5$ (for other
cases see Ref.~\cite{KerenPhysicaB00a}). There are two important features in
this figure. First, at low frequencies there are pronounced wiggles in the
simulation waveform at early times, as in the data. The time scale on which
wiggles are observed [$t_{w}$], which is demonstrated on the graph, obeys $%
t_{w}\lesssim \tau _{l}$. Therefore, as mentioned before, $\tau _{l}$ could
be estimated by visually inspecting the low field early time data. Second,
there is absolutely no difference in the polarization between $\omega =0$
and $\omega =0.3$. In other words, at low fields the polarization is field
independent and the field impacts the polarization only for $\omega >1/\tau
_{l}$. This provides a complementary qualitative method of estimating $\tau
_{l}$. In fact, both methods are well known for the pure exponential
correlation function. The new point being made here is that these methods
are correct even when a power law multiplies the exponential decay term in
the expression for $q(t)$, so the direct observation of the threshold value
of the field dependence of the polarization is a complementary,
model-independent method of estimating the cutoff time.

\begin{figure}
\unitlength=1mm
\begin{picture}(100,70)
\put(5,0){\epsfig{file=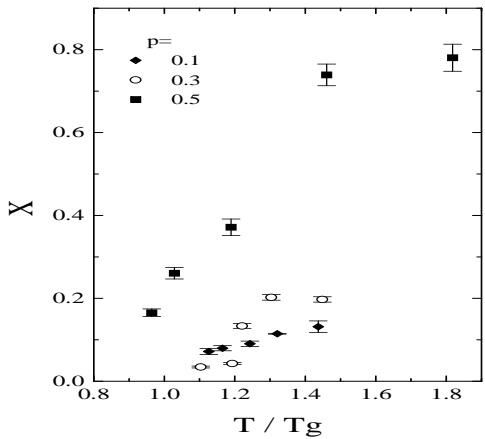,height=70mm,width=80mm}}
\end{picture}
\caption{$x$ (from Eq.~\ref{Itao}) plotted vs. temperature for different
samples.}
\label{XVerT}
\end{figure}

In Fig.~\ref{Numerics}b we show the numerical $P(\omega ,t)$ vs. $t/\omega
^{0.7}$ for $\omega \geq 5$. Clearly, at high enough frequencies and late
enough times scaling of the form of Eq.~\ref{ScalingRelations} indeed holds.
The critical value above which $H$ and $t$ are considered large is
determined by $\tau _{l}$. Late time is $t>\tau _{l}$, and high fields are $%
H>(\gamma _{\mu }\tau _{l})^{-1}$. It is therefore very important to observe
wiggles and/or strong field effect in the data (e.g. Fig.~\ref{Fitting})
before the cutoff time and $x$ can be obtained unambiguously from the
scaling.

We now apply these concepts to the analysis of our $\mu $SR data. The
analysis of the measurements shown in Fig. \ref{Fitting}(a) is discussed in
detail using four steps. First, we estimate $\tau _{l}$. The wiggles in the
waveform prevail for $t_{w}\sim 2$~$\mu $sec, and fields as small as $30$ G (%
$\omega =2.5$ MHz) impact the polarization. Both these observations indicate
that $\tau _{l}$ is of the order of $1$~$\mu $sec. Second, we evaluate $x$.
\ From the value of $\tau _{l}$ we expect scaling to hold for $t>1$~$\mu $%
sec and for fields bigger than $100$~G. Indeed, in Fig.~\ref{Scaling} when $%
t/H^{1}>$ $0.01$ (and even much earlier) all data sets collapse onto one
line, meaning $x\rightarrow 0$ (as mentioned before). Third, we estimate $%
C^{\ast }$ and $\beta $. By expanding $\iota (t)$ in powers of $t$,
performing the integral of Eq.~\ref{DefofGamma} term by term for the zero
field case ($H=0$), and keeping only the first term we find that 
\[
\Gamma ^{\ast }(0,t)t=\frac{2C^{\ast }t^{2-x}}{(2-x)(1-x)}+O(t^{3-x}). 
\]
Using this relation in Eq.~\ref{FinalPolarization}, and fitting the lowest
field early time data to a stretched exponential (while holding $x$ fixed
from second step) we can obtain $C^{\ast }$ and $\beta $. The values found
up to now for $C^{\ast }$, $x$, $\tau _{l}$, and $\beta $ are used as
initial guesses for a global fit (at a given $T$) for all applied fields. In
the global fit we allow freedom in $H$ for fields $\leq $ 30~G in order to
account for the sample's susceptibility, demagnetization, and also since our
power supply does not give accurate values in such small fields. The global
fit produces values of $x$ somewhat larger than those obtained from scaling
and smaller error bars. The results of this fit procedure are presented in
Fig.~\ref{Fitting}(a) by the solid lines. Our model captures the essence of
the wiggles, although for some fields it fails to capture the fine details
of the waveform at early times. More importantly, the model accounts very
well for the data past the wiggles for more than 2.5 orders of magnitude in
field. The same analysis method is applied to our series of samples at
various temperatures. We find that $\beta =0.45(5)$ globally for all samples
and temperatures. A second example of the success of this analysis procedure
is depicted in Fig.~\ref{Fitting}(b) by the solid lines.

At first sight it would appear that with so many fit parameters no
significant conclusions could be reached. In fact, because a single set of
fit parameters are being used to fit a whole family of curves and because
different parameters dominate the behavior at different ranges of time and
field, the fit parameters are well pinned down. There is some play-off
between different parameters so the values are correlated, but it turns out
that the conclusions that we will reach (relaxation dominated by $x$, cutoff
times comparable or longer than a microsecond) are robust. To get fully
reliable and unique fits, certain purely experimental parameters (background
and initial anisotropy for the whole range of fields used) must be well
established from independent control runs.

\section{Results}

\label{Results}

We now turn to discuss the outcome of the fits. The power $x$ is plotted vs. 
$T/T_{g}$ in Fig.~\ref{XVerT}. At all temperatures studied (from $T_{g}$ to
about $2T_{g}$) we find $x$ values that are clearly bigger than zero, and
that increase with increasing $T$, in a qualitative agreement with
Ogielski's simulations \cite{OgielskiPRB85} and with the more recent results
of Franzese and Coniglio \cite{FranzesePRE99}. In general, at a given $%
T/T_{g}$, the power $x$ appears to increase with increasing magnetic ion
concentration $p$. Theoretically, near a continuous phase transition the
power $x$ is expected to relate to the static and dynamic critical exponents
through $x=(d+\eta )/2z$ \cite{OgielskiPRB85}. We find $x$ values which are
of the order of $0.15$ close to $T_{g}$. This should be compared with the
value of $0.13$ that can be estimated from exponents obtained by magnetic
measurements at low frequencies on the same type of alloy \cite{levy}. This
exponent could be expected to be independent of concentration for dilute
alloys.

\begin{figure}
\unitlength=1mm
\begin{picture}(150,95)
\put(5,0){\epsfig{file=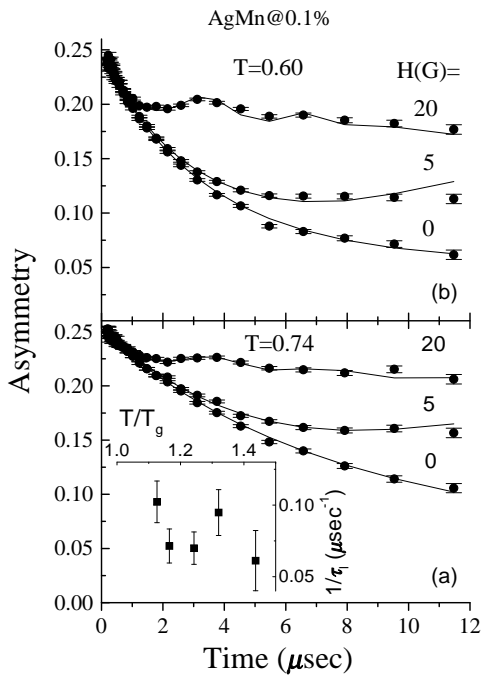,height=100mm,width=70mm}}
\end{picture}
\caption{Demonstrating that neither the wiggling time $t_{w}$, nor the
sensitivity of the polarization to the field are varying with temperature.
The inset shows the late cutoff time $\protect\tau _{l}$ as given by Eq.~\ref
{Itao} versus $T/T_g$ for the $p=0.1$ sample.}
\label{taulVerT}
\end{figure}

As for the cutoff time $\tau _{l}$, we have noticed that $\tau _{l}$ changes
little as a function of the temperature over the temperature range we have
studied. This is best demonstrated in the $p=0.1$ sample where the waveform
has strongest wiggles and strong field dependence, and is therefore most
sensitive to $\tau _{l}$. In Fig.~\ref{taulVerT}(a) and (b) we present $%
P(H,t)$ at three representative fields and two different temperatures
approaching $T_{g}$. Clearly a field as small as $5$ G ($\omega =0.4$ MHz)
impacts the polarization at both temperatures, and there is no big change in
the wiggle time $t_{w}$ ($\sim 10$ $\mu $sec) between these temperatures.
Thus $\tau _{l}$ remains longer than a few microseconds up to $T/T_{g}=1.43$%
. In the inset of Fig.~\ref{taulVerT}(a) we show the fit results for $1/\tau
_{l}$ for the $p=0.1$ sample, demonstrating once again that we do not
observe a critical temperature dependence. On general physical principles
related to any continuous transition, $\tau _{l}$ should diverge as $T$
tends to the ordering temperature (see for instance Ogielski's Ising spin
glass simulations \cite{OgielskiPRB85}). The present data simply indicate
that for the concentrations and temperature ranges studied, $\tau _{l}$ is
at the edge or out of the muon dynamic window.

\begin{figure}
\unitlength=1mm
\begin{picture}(100,70)
\put(5,0){\epsfig{file=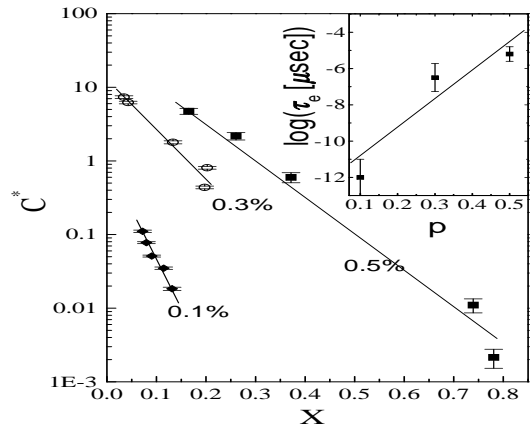,height=70mm,width=80mm}}
\end{picture}
\caption{The prefactor $C^{\ast }$ plotted versus $x$ on a semi-log scale
(see Eq.~\ref{CStar}). The temperature is an implicit parameter. The solid
curves are linear fits. Their slope is interpretted as $\log \protect\tau
_{e}$ and presented in the inset. }
\label{cVerx}
\end{figure}

Finally, we examine the parameter $C^{\ast }$. The mean square of the field $%
\left\langle {\bf B}_{\bot }^{2}\right\rangle $ (or $\Delta ^{2}$) at a
given site should be $T$ independent, and so should $(\Delta ^{\ast })^{2}$.
If $\tau _{e}$ is also $T$ independent, we expect from Eq.~\ref{CStar} a $%
\log (C^{\ast })$ against $x$ plot to be a straight line. Indeed, a linear
dependency between $\log (C^{\ast })$ and $x$ is consistent with the data as
shown in Fig.~\ref{cVerx}. The slope of the line at each concentration is $%
\log \tau _{e}$ and the intercept is $2\log (\Delta ^{\ast })$. The values
of the intercepts have large error bars (since we do not have data in
sufficiently small $x$) and are therefore not presented. Nevertheless, a
rough estimate of $\Delta ^{\ast }$ indicates that the condition for
validity of our analysis, namely, $H>\Delta ^{\ast }$ (see appendix) holds
for $H\gtrsim 10$~G for the $p=0.5$ and $0.3$ samples, and $H\gtrsim 1$~G
for the $p>0.1$ sample\ as done here. The values of $\log \tau _{e}$, are
presented in the inset of Fig.~\ref{cVerx} for the different samples. These
values make the conjecture $t\gg \tau _{e}$ self-consistent, although we
find $\tau _{e}$ for the $p=0.1$ sample unacceptably small. Nevertheless,
our analysis indicates that the $T$ dependence of $C^{\ast }$ is due only to 
$x(T)$.

To summarize, we demonstrate that a distribution of correlation times cannot
account for the muon polarization as a function of both field and time.
However, we can account for our data provided that each muon experience a
local field correlation function given by Eq.~\ref{PartDefOfq}, where the
only difference between different muons is in the value of $\Delta $. Our
analysis shows that out of all the parameters that determine $P(H,t)$,
namely, $x$, $\tau _{l}$, $\tau _{e}$, and $\Delta ^{\ast }$, the only one
that is temperature dependent is $x$. We thus conclude that the dynamic
behavior of this family of dilute spin glasses near $T_{g}$ is controlled
essentially by the temperature dependence of the power law exponent $x$.

\section{Acknowledgments}

We would like to thank the ISIS facility for its hospitality. This research
was supported by the Israeli Academy of Science, and by the EU TMR program.
We also would like to thank C. Paulsen for measuring $T_{g}$ in our low
concentration samples using his dilution refrigerator.

\bigskip \appendix

\section{From field correlations to spin lattice relaxation}

\label{Appendix}

In this appendix we derive the expected behavior of the polarization of a
local probe, which is out of equilibrium, in a dynamic field environment.
For simplicity we use a semi-classical approach. A more complete result,
based on a full quantum treatment, can be found in the literature \cite
{McMullenPRB78} but there is no detectable difference. In the semi-classical
method we decompose the Hamiltonian of a spin $1/2$ experiencing both an
external static field $H$, and an internal dynamically fluctuating field $%
{\bf B}^{d}(t)$ into 
\begin{equation}
{\cal H}={\cal H}_{0}+{\cal H^{\prime }}(t)
\end{equation}
where 
\begin{equation}
{\cal H}_{0}=-\gamma _{\mu }S_{z}H  \label{MusrZeroHamil}
\end{equation}
is the secular part, and the interaction part is 
\begin{equation}
{\cal H^{\prime }}=-\gamma _{\mu }{\bf S}\cdot {\bf B}^{d}(t).
\end{equation}
The time dependent field ${\bf B}^{d}(t)$ is taken to be classical and ${\bf %
S}$ is the muon spin operator. When the fluctuating fields are smaller than
the external field we can use time dependent perturbation theory and write
the time propagator as 
\begin{eqnarray}
U(t) &=&\exp (-\frac{i}{\hbar }{\cal H}_{0}t)[1-\frac{i}{h}%
\int_{0}^{t}dt^{\prime }{\cal H}^{\text{I}}(t^{\prime })-
\label{QuantTimePropaget} \\
&&\frac{1}{h^{2}}\int_{0}^{t}dt^{\prime }\int_{0}^{t^{\prime }}dt^{\prime
\prime }{\cal H}^{\text{I}}(t^{\prime }){\cal H}^{\text{I}}(t^{\prime \prime
})+\ldots ].  \nonumber
\end{eqnarray}
where the perturbation Hamiltonian in the interaction picture is given by 
\begin{equation}
{\cal H}^{\text{I}}(t)=\exp (i{\cal H}_{0}t/\hbar ){\cal H}^{\prime }(t)\exp
(-i{\cal H}_{0}t/\hbar ).
\end{equation}
This Hamiltonian simplifies to 
\begin{equation}
{\cal H}^{\text{I}}(t)=-\gamma _{\mu }{\bf B}^{d}(t){\bf S}^{\text{I}}(t)
\label{IntHamiltoinan}
\end{equation}
where 
\[
{\bf S}^{\text{I}}(t)=\exp (i\omega S_{z}t/\hbar ){\bf S}\exp (-i\omega
S_{z}t/\hbar ) 
\]
and 
\[
\omega =-\gamma _{\mu }H. 
\]
Explicitly ${\bf S}^{\text{I}}(t)$ is given by 
\begin{equation}
\begin{array}{lll}
S_{x}^{\text{I}}(t) & = & S_{x}\cos (\omega t)-S_{y}\sin (\omega t) \\ 
S_{y}^{\text{I}}(t) & = & S_{y}\cos (\omega t)+S_{x}\sin (\omega t) \\ 
S_{z}^{\text{I}}(t) & = & S_{z}.
\end{array}
\label{SpinOperator}
\end{equation}
Note that ${\bf S}^{\text{I}}(t)$ is the time-dependent spin operator in the
interaction picture, namely, of a muon that rotates around the external
field as if there were no internal fields. It is not the spin operator ${\bf %
S}(t)$ of a muon that experiences the combined static and dynamic fields. In
other words, in order to find ${\bf S}(t)$ we use a perturbation expansion
in terms of ${\bf S}^{\text{I}}(t)$. This is the (mathematical) origin of
the wiggles seen in the data at the frequency of the external field.

Using Eqs.~\ref{IntHamiltoinan} and \ref{SpinOperator} we find that 
\begin{equation}
{\cal H}^{\text{I}}(t)=-\gamma _{\mu }\left[ V(t)S_{x}-U(t)S_{y}+T(t)S_{z}%
\right]  \label{MusrExpIntHam}
\end{equation}
where 
\[
\begin{array}{lll}
V(t) & = & B_{x}^{d}(t)\cos (\omega t)+B_{y}^{d}(t)\sin (\omega t) \\ 
U(t) & = & B_{x}^{d}(t)\sin (\omega t)-B_{y}^{d}(t)\cos (\omega t) \\ 
T(t) & = & B_{z}^{d}(t).
\end{array}
\]
The polarization of a muon at a given site as a function of time $P_{z}(t)$
is given by 
\begin{equation}
P_{z}(t)=\text{Tr}\left[ \rho U^{\dagger }(t)\sigma _{z}U(t)\right]
\label{PolarizExpr}
\end{equation}
where 
\[
\rho =\frac{1+P_{0}\sigma _{z}}{2} 
\]
and $P_{0}$ is the initial polarization.

Equations~\ref{PolarizExpr} and \ref{QuantTimePropaget} lead to the
perturbation series 
\begin{equation}
P(t)=I+II+III+...
\end{equation}
where 
\begin{equation}
\begin{array}{lll}
I & = & \text{Tr}\left\{ \rho \sigma _{z}\right\} , \\ 
II & = & \frac{i}{\hbar }\int_{0}^{t}dt^{\prime }\text{Tr}\rho \lbrack {\cal %
H}^{\text{I}}(t^{\prime }),\sigma _{z}], \\ 
III & = & -\frac{1}{\hbar ^{2}}\int_{0}^{t}dt^{\prime }\int_{0}^{t^{\prime
}}dt^{\prime \prime }\text{Tr}\rho \lbrack {\cal H}^{\text{I}}(t^{\prime
\prime }),[{\cal H}^{\text{I}}(t^{\prime }),\sigma _{z}]].
\end{array}
.  \label{Terms}
\end{equation}
We now evaluate each term explicitly: The first term is simple. Since 
\[
\text{Tr}\left\{ \rho \sigma _{i}\right\} =\left\{ 
\begin{array}{cc}
P_{0} & i=z \\ 
0 & i=x,y
\end{array}
\right| 
\]
we get $I=P_{0}$. In the second term we have 
\[
\lbrack {\cal H}^{\text{I}}(t^{\prime }),\sigma _{z}]=i\gamma _{\mu }\left(
V(t^{\prime })\sigma _{y}+U(t^{\prime })\sigma _{x}\right) . 
\]
After the evaluation of the trace we find $II=0$. For the third term we find 
\[
\text{Tr}\rho \lbrack {\cal H}^{\prime }(t^{\prime \prime }),[{\cal H}%
^{\prime }(t^{\prime }),\sigma _{z}]]=P_{0}\gamma _{\mu }^{2}\left[
V(t^{\prime \prime })V(t^{\prime })+U(t^{\prime \prime })U(t^{\prime })%
\right] 
\]
where we have used the simplifying assumption that $V$, $U$, and $T$ are
classical fields. To further progress with the calculation it is helpful to
introduce at this stage the ensemble average over all possible trajectories
of ${\bf B}_{\bot }^{d}(t).$ Quantities in $\left\langle {}\right\rangle $
will denote such an average and we can write 
\begin{equation}
\left\langle P_{z}\right\rangle =\left\langle I\right\rangle +\left\langle
III\right\rangle  \label{PExpantion}
\end{equation}
where $\left\langle I\right\rangle =P_{0}$ and 
\[
\left\langle III\right\rangle =P_{0}\gamma _{\mu }^{2}\left\langle
V(t^{\prime \prime })V(t^{\prime })+U(t^{\prime \prime })U(t^{\prime
})\right\rangle . 
\]
The meaning $\left\langle B_{i}^{d}(t^{\prime })B_{j}^{d}(t^{\prime \prime
})\right\rangle $ is to hold $t^{\prime }$ and $t^{\prime \prime }$ fixed
and to average over all possible fields in the systems in between these
times. This is a correlation function. We assume that there are no cross
correlations, and x,y symmetry so the only terms left to evaluate are of the
form 
\begin{eqnarray*}
&&\left\langle V(t^{\prime \prime })V(t^{\prime })+U(t^{\prime \prime
})U(t^{\prime })\right\rangle  \\
&=&\left\langle B_{x}^{d}(t^{\prime })B_{x}^{d}(t^{\prime \prime
})+B_{y}^{d}(t^{\prime })B_{y}^{d}(t^{\prime \prime })\right\rangle \cos
\left( \omega (t^{\prime \prime }-t^{\prime })\right) .
\end{eqnarray*}
We now make the assumption that the correlation function depends only on the
time difference and define 
\begin{equation}
\Phi (t^{\prime }-t^{\prime \prime })=\gamma _{\mu }^{2}\left\langle
B_{x}^{d}(t^{\prime })B_{x}^{d}(t^{\prime \prime })+B_{y}^{d}(t^{\prime
})B_{y}^{d}(t^{\prime \prime })\right\rangle .
\end{equation}
so that finally 
\[
\left\langle III\right\rangle =-P_{0}\int_{0}^{t}dt^{\prime
}\int_{0}^{t^{\prime }}dt^{\prime \prime }\Phi (t^{\prime }-t^{\prime \prime
})\cos \left( \omega (t^{\prime }-t^{\prime \prime })\right) . 
\]
We can now eliminate one of the integrals by writing 
\[
\begin{array}{lll}
\tau & = & t^{\prime }-t^{\prime \prime } \\ 
\tau ^{\prime } & = & t-t^{\prime }
\end{array}
. 
\]
This transforms the integral into 
\[
\int_{0}^{t}dt^{\prime }\int_{0}^{t^{\prime }}dt^{\prime \prime
}=\int_{0}^{t}d\tau \int_{0}^{t-\tau }d\tau ^{\prime } 
\]
and we arrive at 
\begin{equation}
\left\langle III\right\rangle =-P_{0}\int_{0}^{t}d\tau (t-\tau )\Phi (\tau
)\cos (\omega \tau ).  \label{FianlIII}
\end{equation}

If we now write 
\begin{equation}
P_{z}(t)=P_{0}\exp \left( -\Gamma (t)t\right) ,
\end{equation}
expand this equation in powers of $\Gamma $, and compare it with Eqs.~\ref
{PExpantion} and \ref{FianlIII} we find 
\[
\Gamma (t)t=\int_{0}^{t}d\tau (t-\tau )\Phi (\tau )\cos (\omega \tau ).
\]
At {\em late times}, such that $\Phi (t)$ is negligible, one finds that 
\begin{equation}
\Gamma (H,t)\rightarrow \frac{1}{T_{1}(H)}=\int_{0}^{\infty }\Phi (\tau
)\cos (\omega \tau )d\tau .  \label{SpectralDensity}
\end{equation}

\end{document}